# The alternative fact of "probable vaccine damage":
# A typology of vaccination beliefs in 28 European countries

Simona - Nicoleta Vulpe[a] and Cosima Rughiniș[b]

[a] Interdisciplinary School of Doctoral Studies, University of Bucharest, 36-46 Mihail Kogălniceanu, Bucharest, Romania; mona.vulpe@gmail.com

[b] Department of Sociology, Faculty of Sociology and Social Work, University of Bucharest, 9 Schitu Măgureanu, Bucharest, Romania; cosima.rughnis@gmail.com

**Corresponding author:**

Simona - Nicoleta Vulpe, 36-46 Mihail Kogălniceanu, room 221 first floor, Sector 5, Bucharest, 050107, Romania. E-mail: mona.vulpe@gmail.com. Phone number: +40 742439983

# The alternative fact of "probable vaccine damage":

# A typology of vaccination beliefs in 28 European countries

**Abstract**

*Background:* Despite lacking scientific support, vaccine hesitancy is widespread. While "vaccine damage" as a scientific fact is statistically highly uncommon, emerging social and technological forces have converted "probable vaccine damage" into an alternative fact, thus making it a widely shared intersubjective reality.

*Methods:* Using the Eurobarometer 91.2 survey on a statistically representative EU27-UK sample interviewed in March 2019, we documented perceptions of vaccine risks and identified three belief configurations regarding vaccine effectiveness, safety, and usefulness, through exploratory cluster analysis.

*Results:* The public beliefs in vaccine risks are frequent. Approximatively one-tenth of the EU27-UK population consider that vaccines are not rigorously tested before authorization, one-third believe that vaccines can overload or weaken the immune system and that they can cause the disease against which they protect, and almost one-half believe that vaccines can cause serious side effects. We identified three belief configurations: the skeptical, the confident, and the trade-off clusters. The skeptical type (approx. 11% of EU27-UK respondents) is defined by the belief that vaccines are rather ineffective, affected by risks of probable vaccine damage, not well-tested, and useless; the confident type (approx. 59%) is defined by beliefs that vaccines are effective, safe, well-tested, and useful; and the trade-off type (approx. 29%) combines beliefs that vaccines are effective, well-tested and useful, with beliefs of probable vaccine damage. The vaccine-confident and the trade-off types profess having similar vaccination histories, indicating the significant role of other factors besides beliefs in inducing behavior.

*Conclusions:* The high proportion of varying public beliefs in vaccine risks and the presence of a trade-off type of vaccination assessment indicate the social normality of beliefs in probable vaccine damage. Probable vaccine damage presently exists as an alternative fact in the public imagination, perceptively available for wide segments of the public, including those who trust medical science.

## Introduction

Vaccine hesitancy is one of the main global health threats, according to the World Health Organization (WHO) [1]. In recent years, measles epidemics threatening public health in the United States and in European countries were caused by low rates of vaccination. In the future, SARS-CoV-2 vaccine might be met with resistance from vaccine-skeptical people, needlessly prolonging the pandemic. Thus, in this paper, we document the current distribution and configurations of vaccine-skeptical beliefs in the European Union (EU) and the United Kingdom (UK) and their relationship with vaccination behavior and sociodemographic categories, and we discuss the significance of our findings concerning vaccination campaigns.

The WHO defines vaccine hesitancy as low vaccine uptake under circumstances of availability of vaccination services [2], combining dimensions related to behavior, attitude, and belief. Vaccine hesitancy can be understood as attitudes comprising "varying degrees of indecision" [3] justified through beliefs about vaccine risks and benefits, which motivate people to selectively accept, delay, or reject vaccines [4].

Historically, vaccine hesitancy can be said to be older than vaccines themselves, since it originated with the practice of variolation that preceded vaccines and vaccination [5]. Resistance and opposition to vaccination are shaped, in intensity and distribution, by the social organization of a given epoch and society [6]. In the late nineteenth and early twentieth centuries, religious worldviews were strongly related to anti-vaccination and to the creation and propagation of anti-scientific discourses [7]. Other factors that were shaping early anti-vaccination movements were the risks carried by inoculation and vaccination, as well as political opposition to state intervention. The forces that shape anti-vaccination movements and discourses today are quite different, resulting from novel ideological, economic, and technological drives [8,9].

Dramatic successes in fighting disease and decreases in vaccine-related adverse effects have not led to unanimous trust in science and biomedicine since they have been accompanied by structural changes that have kindled novel forms of vaccine hesitancy. Increased demand for individual responsibility triggered by reflexive modernity and neoliberal capitalism [10,11] have placed people in the emerging role of the "informed patient" [12,13] expected to make autonomous judgments and to balance conflicting health-related assessments. In a context of media amplification of vaccine scares [14] and an online environment densely interlinked with accounts of vaccine damage [15,16], the results of these judgments have often led to mistrust of medical authority and science.

Present-day vaccine hesitancy has been shaped by a recent history of vaccine scares that raised the issue of trust in vaccination on the public agenda, including the UK pertussis controversy, the mercury poisoning controversy, and the alleged measles, mumps, rubella (MMR)–autism vaccination link [17,18]. Social anxieties related to the immune system and the very centrality of it in public discourses on health were amplified by the AIDS/HIV crisis and by the difficult to explain rise of autoimmune disorders in developed countries [18].

At the same time, the internet and, particularly, Web 2.0 with its wealth of user-generated content, social networks, and digital communities have boosted the visibility and diversity of vaccine-skeptical messages [5,18,20]. A complex information landscape emerged that patients and practitioners alike now have to navigate [15,16], from individual and organized testimonies of vaccine damage [20] to pseudo-scientific theories, and from alternative views of health and well-being derived from complementary and alternative medicine [7,20] to distorted representations of scientific research and even the propagation of fraudulent studies. People from all walks of life and all levels of expertise have learned to make their way in this new informational landscape, participating agentively and creatively in novel forms of folklore [22] and "urban myths" [23], making vaccination decisions based on many factors. In this sensemaking effort, they are displaying variable cognitive styles [24] and divergent popular epistemologies [25–27], and they are developing reflexive lay methods of searching for, curating, and interpreting information [12,28]. The popularity of vaccine-skeptical beliefs and opinions and their diversity has gradually increased in this emerging social configuration, in which scientists and medical practitioners are only a few of the voices to be heard and trusted. Vaccine-confident and vaccine-skeptical people engage in mutual and reciprocal negative stereotyping as "subject to epistemic closure, groupthink, confirmation bias and over-confidence in their own expertise" [21,29], rendering dialogue difficult.

Vaccine hesitancy can be conceptualized and measured on a continuum, from low to high. At the same time, it is a multidimensional phenomenon, which depends on perceptions of vaccine risks, benefits, effectiveness, or usefulness at the social level, as well as on the dimensions that capture people's overall assessments of risk and benefits, thus generating attitudes towards vaccination, in general, and specific vaccines, in particular, and intentions to vaccinate completely and in due time, selectively or with modified schedules, or not at all. A closer look at vaccine hesitancy in a population will identify diverse configurations, types of beliefs, and attitudes, which better describe people's position in regard to vaccination than unidimensional estimates. Vaccination perception can be understood through Doty's metaphor of a "buffet of beliefs" [28], in which people combine what makes sense to them in multifarious configurations.

One interesting and worrying development consists of the emergence of widespread perceptions of probable vaccine-related harms. The scientific consensus is that serious vaccine adverse effects are uncommon, occurring with probabilities that are degrees of magnitude lower than the harms that would be incurred through vaccine preventable diseases. Still, the public perception of vaccine-related adverse effects does not reflect this consensus [16,31]. In recent years, this perception of "probable vaccine damage" has become an alternative fact, easily available to those engaged in information seeking, especially from online venues [26,32]. Research has systematically documented large proportions of populations who believe that vaccines pose significant health risks. The popularity of these beliefs becomes self-reinforcing through processes of social construction of reality, as people contribute in the co-creation of shared worldviews that are amplified in social bubbles. Belief in vaccination harm can coexist in diverse configurations of vaccination hesitancy [28], as it can be accompanied by beliefs in vaccination effectiveness and usefulness, or generalized rejection of vaccination.

There are stable dominant themes in people's beliefs about the adverse effects of vaccines [32]. In a systematic review of 71 studies of beliefs about vaccines in the US, with a focus on barriers to vaccination among minority populations, the authors found that "The most frequently cited beliefs included that vaccines can cause illnesses; a child's immune system can be overwhelmed if receiving too many vaccines at a time; vaccines contain harmful ingredients; younger children are more susceptible to vaccine adverse events; the purpose of vaccines is profit-making; and naturally developed immunity is better than that acquired from vaccines" [34, p6]. Vaccines as a cause of illnesses is a wide-ranging theme within the vaccine hesitant discourse. It goes beyond autism as a vaccine-induced condition and lists "dysfunction of the immune system, developmental and neurological disorders, behavioral issues, diabetes, liver problems, cancer, and death" (34, p6] as side effects of vaccines. The "overwhelming" of the immune system and the toxicity of vaccine ingredients and mercury, in particular, are also recurrent themes within the skeptical discourses on vaccines [35,36].

Multiple studies in diverse societies have indicated that beliefs in vaccine adverse effects and concerns about vaccine safety are widespread, both for people who vaccinate and those who are hesitant, throughout the globe [36]. A 2010 survey of US parents of children six years of age or younger found that significant proportions of parents expressed safety concerns such as: "Vaccines may cause learning disabilities, such as autism" (30%), "The ingredients in vaccines are unsafe" (26%), "Vaccines are not tested enough for safety" (17%), and "Vaccines may cause chronic disease" (16%) [37]. These concerns were also frequently found among parents who had fully vaccinated their children or intended to do so, even if less frequently than among vaccine hesitant parents [38]. In 2010, a study of European parents found that a significant minority "have ever worried about the safety of a vaccination," ranging from 4% in Norway to 21% in Spain [38, p. 5735]. A 2012 survey of vaccine knowledge in Switzerland identified several misconceptions that were frequent among the respondents: vaccination may overload the immune system (30%), vaccinations are administered too early and prevent the development of the immune system (40%), vaccines include chemical ingredients in doses dangerous for humans (37%), vaccines lead to increased occurrence of allergies (35%), and vaccines may trigger autism, multiple sclerosis, and diabetes (16%) [39]. A 2016 survey in France identified 26% of parents as "vaccine refusers," 7% as "delayers," and another 13% as "acceptors with doubts" [40]. In a 2016 survey in Australia, the authors found overlapping trust and mistrust in vaccination: "The vast majority of parents were supportive of vaccination in children: 68% strongly support, 26% generally support, 2% are neutral, 2% generally oppose and 2% strongly oppose. (…) While 90% of parents agreed that vaccinations were safe for children, 23% were concerned that vaccines were not tested enough for safety, 21% believed that vaccines could cause autism and 22% were also concerned that their child's immune system could be weakened by vaccinations" [41, p. 146]. In 2017, approximately 63% of Italian parents reported some worries with regard to vaccines' side effects, while another 28% were very worried about vaccine safety [42]. In a 2018 survey of parents of young children in 18 European countries, 77% of respondents agreed with the statement, "Overall, I think vaccines are safe," while 16% were not sure, and 7% disagreed [9].

Vaccine-skeptical people may generally display different cognitive styles [24] and ideological commitments [22] than people who trust vaccination. Vaccine skepticism is associated with biased risk processing [43] and a specific psychological pattern comprising higher conspiratorial thinking, reactance, disgust towards blood and needles, and individualistic/hierarchical worldviews [20,22]. At the same time, differences in sociodemographic characteristics are not very pronounced [22], as vaccination skepticism permeates all social strata and categories. Regarding beliefs on vaccine safety, a global survey across 67 countries found that perceptions of vaccine risks do not correlate with gender or education, while being stronger for younger people, the unemployed, and those in the lowest income quartile [36].

Typological studies assessing the diversity in patterns of vaccine hesitancy are relatively rare compared with linear estimates of intensity and factors, but they systematically indicate the presence of intermediate, combined configurations in which forms of trust and mistrust of vaccination coexist. An example of this is the following classification of US parents: "vaccine believer" (high level of confidence in vaccine safety), "cautious parents" (feel distress when their children are vaccinated because of emotional investment), "relaxed parents" (hold slightly negative views on vaccines), and "unconvinced parents" (hold the most negative views on vaccines) [44]. A cluster analysis of US parents who rejected HPV vaccination resulted in five groups of reasons for vaccine rejection: "pragmatic concerns about effects on sexual behavior, specific HPV vaccine concerns, moral concerns about sexual behavior, general vaccine concerns, and denial of need" [45, p108]. Three clusters of US people rejecting vaccination against A/H1N1 were identified while examining their potential for attitude change: "open to persuasion," "informed unconvinced," and "disengaged skeptics" [46]. Using latent profile analysis, three types for Australian parents active on social media were found: "accepters" (vaccine confident, intend to vaccinate), "fence sitters" (believe in benefits of vaccines, reject mandatory vaccination), and "rejecters" (reject all vaccines) [47].

Based on our proposed empirical typology of respondents in the EU and the UK, we will document how perceptions of vaccination harm are not specific to a distinctive vaccine-skeptical worldview. On the contrary, beliefs that vaccination is risky have become widespread and mainstream, making vaccine damage and vaccination risks into plausible, probable, and feared events for large segments of the population.

**Methods**

We devised an exploratory classification of people's beliefs about vaccine effectiveness, risks, and usefulness among the European population. The data set was sourced from *Eurobarometer 91.2* [48]. The weighted sample was representative for the EU27-UK. The Eurobarometer sample was a multistage, random probability sample consisting of 27,524 respondents 15 years of age and older.

We conducted a K-means cluster analysis using IBM SPSS Statistics version 23. We selected a three cluster classification that we considered would be able to capture the distinctive and interpretable configurations at the level of the EU27-UK sample. Assuming missing completely at random (MCAR) data and given the low percentage of missing observations (from 2.4% to 4.5% for cluster indicator variables), pairwise deletion was used as a method to handle missing data.

The variables included in our cluster analysis measured perceived effectiveness of vaccines, beliefs about vaccination risks, and beliefs about usefulness of vaccination. Indicators for each variable were recoded into dichotomous indicators to facilitate comparability and cluster interpretability.

The perceived effectiveness of vaccines was measured by the following indicator: "All the diseases mentioned earlier are infectious diseases and can be prevented. Do you think that vaccines can be effective in preventing them?" Response options included: "Yes, definitely," "Yes, probably," "No, probably not," and "No, not at all," and they were dichotomized into "Yes" (definitely or probably) or "No" (definitely or probably).

Beliefs about vaccination risks were elicited using the following prompt: "For each of the following statements, could you please tell me whether you think it is true or false…." Four items were presented with the dichotomous response options of "True" and "False," for which we marked in parentheses the scientifically correct one: "Vaccines overload and weaken the immune system" (False), "Vaccines can cause the disease against which they protect" (False), "Vaccines can often produce serious side-effects" (False), and "Vaccines are rigorously tested before being authorized for use" (True). We dichotomized variables by singling out responses that diverged from the scientific consensus and grouping uncertainty (Don't know) with the scientifically correct responses.

Beliefs about usefulness of vaccination was measured by asking: "To what extent do you agree or disagree with the following statements?" and listing the following indicators: "It is important for everybody to have routine vaccinations," "Vaccines are only important for children," "Not getting vaccinated can lead to serious health issues," "Vaccines are important to protect not only yourself but also others," and "Vaccination of other people is important to protect those that cannot be vaccinated (e.g., newborn children, immunodepressed or very sick people)." Response options included: "Totally agree," "Tend to agree," "Tend to disagree," and "Totally disagree," which we dichotomized into "agree" or "disagree" answers.

# Results

A descriptive analysis of popular beliefs in vaccine safety indicates that large segments of the population in each country consider that vaccine damage is a salient risk (see Figure 1). At the general level of the EU27-UK, approximatively one-tenth of the population consider that vaccines are not rigorously tested before authorization, one-third believe that vaccines can overload or weaken the immune system and that they can cause the disease against which they protect, and almost half of respondents believe that vaccines can cause serious side effects.

*Figure 1. Country distribution of beliefs in vaccine effectiveness, risks and usefulness. Data source: Eurobarometer 91.2*

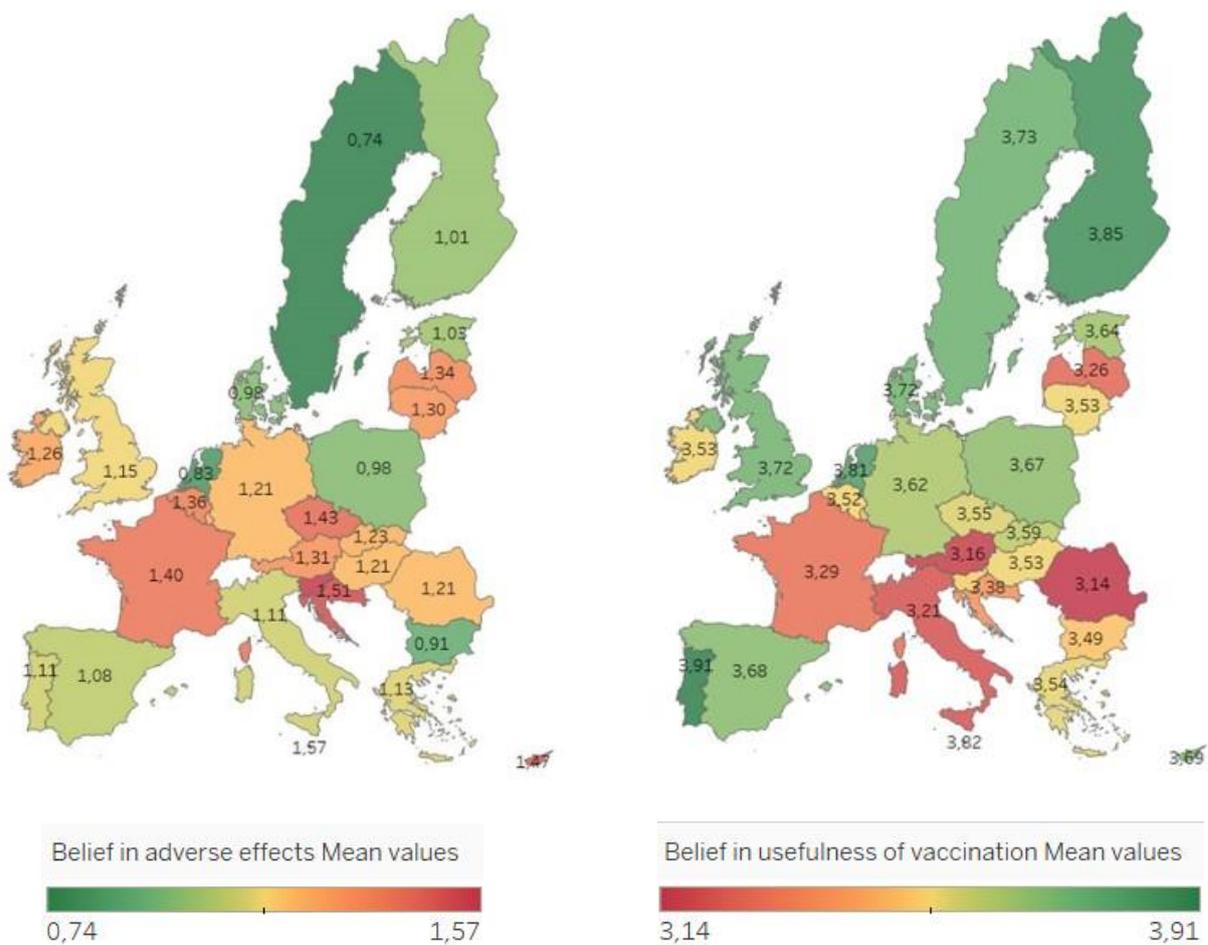

The perception that vaccines are risky is not just the property of people who mistrust vaccination or science. At these levels of frequency, it has become an "alternative fact" [49] *avant la lettre*, available for perception especially by exploring online information for people of all cognitive and ideological styles. In order to document how this alternative fact is distributed in patterns of beliefs regarding vaccination, we classified respondents through K-means cluster typing of their perceptions of vaccine effectiveness, safety, and usefulness.

The exploratory classification yielded three distinctive patterns of beliefs (see Table 1). Skeptical respondents are consistently mistrustful of vaccination, rejecting its benefits and usefulness, and having a high probability of believing in vaccine adverse effects. The confident respondents display the reciprocal pattern of consistent trust in the effectiveness, safety, and usefulness of vaccines. The trade-off type combines high perceptions of vaccination risks with high assessments of vaccination effectiveness and usefulness.

Interestingly, the trade-off cluster had the highest score on all indicators of belief in vaccination harms (for example 91% of the trade-off cluster agreed that "vaccines can often produce serious side-effects" in comparison to 72% of the skeptical and 22% of the confident) with the exception of assessing vaccination testing, on which the skeptical ranked highest in mistrust (41% of the skeptics disagreed that vaccines are rigorously tested before use).

It is also noteworthy that, while respondents classified as the confident type had the lowest rates of belief in vaccination risks, the incidence was not negligible. About 13% of vaccine-confident respondents answered that "vaccine can cause the disease against which they protect," and about 22% answered that "vaccines can produce serious side-effects." Still, only about 5–6% agreed to vaccines having reactions of weakening the immune system or to vaccines not being rigorously tested.

*Table 1. Exploratory classification of EU27-UK respondents according to their belief patterns on vaccine effectiveness, safety and usefulness. Method: K-means cluster. Data source: Eurobarometer 91.2*

| Indicators: | Skeptical | Confident | Trade-off |
|---|---|---|---|
|  | **Mean** | **Mean** | **Mean** |
| Do you think that vaccines can be effective in preventing diseases? *(0 No; 1 Yes)* | **0.39** | 0.95 | **0.90** |

| Indicators: | Skeptical | Confident | Trade-off |
|---|---|---|---|
| Vaccines overload and weaken the immune system *(0 False or DK; 1 True)* | **0.59** | 0.05 | **0.71** |
| Vaccines can cause the disease against which they protect *(0 False or DK; 1 True)* | **0.53** | 0.13 | **0.81** |
| Vaccines can often produce serious side-effects *(0 False or DK; 1 True)* | **0.72** | 0.22 | **0.91** |
| Vaccines are rigorously tested before being authorized for use *(0 True or DK; 1 False)* | **0.41** | 0.06 | 0.10 |
| It is important for everybody to have routine vaccinations *(0 Disagree; 1 Agree)* | 0.15 | 0.95 | 0.90 |
| Vaccines are only important for children *(0 Disagree; 1 Agree)* | 0.46 | 0.24 | 0.34 |
| Not getting vaccinated can lead to serious health issues *(0 Disagree; 1 Agree)* | 0.17 | 0.93 | **0.92** |
| Vaccines are important to protect not only yourself but also others *(0 Disagree; 1 Agree)* | 0.32 | 0.98 | **0.97** |
| Vaccination of other people is important to protect those that cannot be vaccinated (e.g. newborn children, immunodepressed or very sick people) *(0 Disagree; 1 Agree)* | 0.36 | 0.97 | **0.96** |
| **Weighted number of cases** | 3,093 | 16,376 | 8,055 |
| **Percentage within EU27-UK (N=27,524)** | **11%** | **59%** | **29%** |

There is considerable country-level diversity in cluster distribution and in perceptions of probable vaccine damage. In Figure 2, the red/green color scale is used to signal countries with higher/lower risks of vaccine hesitancy, respectively, while a blue scale is used to signal the distribution of the ambivalent trade-off cluster. Some countries are consistently positioned, such as France, Italy, Austria, Latvia, Greece, and Romania, which are on the red spectrum of risk given their relatively high proportions of vaccine-skeptical people and low proportions of vaccine-confident people. At the other end of the continuum, Norway, Sweden, Estonia, Poland, and Spain are consistently in a relatively favorable position for vaccination. Other countries are heterogeneously marked due to their high proportions of the ambivalent trade-off types. The Czech Republic, for example, has a relatively low proportion of vaccine-confident people (51%) but also a low percentage of vaccine-skeptical people (9%), with a high proportion of trade-off respondents that might oscillate between lower or higher vaccine hesitancy. We found Slovakia, Germany, Slovenia, and Lithuania to be in a similar situation.

*Figure 2. Country distribution of vaccine-confident, vaccine-skeptical and trade-off types. Data source: Eurobarometer 91.2*

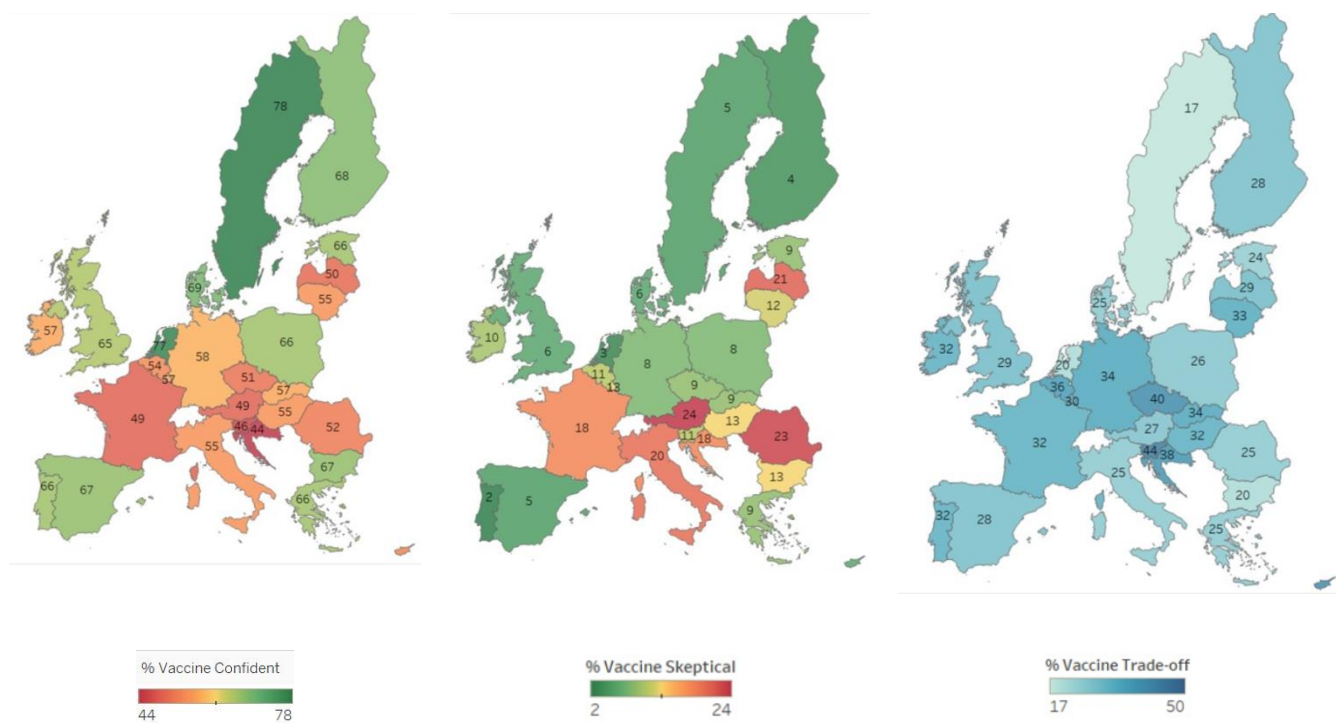

As regards professed vaccination experience, the confident cluster displayed the highest proportion of respondents who either had themselves or their children vaccinated in the last five years, closely followed by the trade-off cluster. We determined a large behavioral gap between the skeptical cluster and the other two types (see Table 2).

*Table 2. Distribution of vaccination behavior across clusters. Data source: Eurobarometer 91.2 (N=27524)*

|  | Skeptical (%) | Confident (%) | Trade-off (%) | Total (%) |
|---|---|---|---|---|
| Has vaccinated her/himself in the last 5 years (for all respondents, N=27524) | 16.1 | 50.2 | 45.4 | 44.9 |
| Has vaccinated children in the last 5 years (for respondents who have children in the household, N=7260) | 52.1 | 75.0 | 67.7 | 70.5 |

Chi-Square tests of association statistically significant for p=0.001

The sociodemographic profiles of the three belief clusters were largely similar. Statistical differences, even when statistically significant (due to the large sample size) were substantively small and did not indicate distinctive social categories. As previous studies [50] have shown, there is a tendency for older generations and, as a consequence, retired people to be more confident in vaccines, a tendency which was also observed in this study. People with higher and ongoing education, as well as students and managers, fall slightly more often in the vaccine-confident category than in the others (see Table 3).

The vaccine-skeptical people positioned themselves more frequently on the right side of an ideological continuum and were also, on average, less satisfied with life than the vaccine-confident and the trade-off types (see Table 3).

*Table 3. Sociodemographic profile of EU27-UK respondents across vaccine belief patterns. Data source: Eurobarometer 91.2 (N=27524)*

|  |  | Skeptical (%) | Confident (%) | Tradeoff (%) |
|---|---|---|---|---|
| Age (years) | 15-24 | 11.2 | 13.2 | 13.3 |
|  | 25-34 | 14.6 | 14.6 | 15.4 |
|  | 35-44 | 18.2 | 15.7 | 16.4 |

|  |  | Skeptical (%) | Confident (%) | Tradeoff (%) |
|---|---|---|---|---|
|  | 45-54 | 18.8 | 16.5 | 17.4 |
|  | 55-64 | 15.8 | 14.4 | 14.8 |
|  | 65–74 | 14.9 | 15.7 | 15.2 |
|  | 75+ | 6.4 | **10.0** | 7.4 |
| Occupation | Self-employed | 7.4 | 7.3 | 6.6 |
|  | Managers | 7.8 | 12.1 | 9.6 |
|  | Other white collars | 13.5 | 11.3 | 12.8 |
|  | Manual workers | 27.5 | 19.2 | 24.6 |
|  | House persons | 4.7 | 5.8 | 5.3 |
|  | Unemployed | 7.3 | 5.6 | 6.2 |
|  | Retired | 24.6 | 28.8 | 26.2 |
|  | Students | 7.3 | 9.9 | 8.7 |
| Education (age when stopped full-time education) | Up to 15 years | 17.5 | 15.6 | 16.0 |
|  | 16–19 years | 44.9 | 39.9 | 46.0 |
|  | 20+ years | 29.4 | 34.0 | 28.8 |
|  | Still studying | 7.4 | 10.0 | 8.8 |
|  | No full-time education | 0.8 | 0.5 | 0.5 |
| Gender | Men | 51.0 | 48.4 | 46.8 |
|  | Women | 49.0 | 51.6 | 53.2 |
| Left-right positioning | 1 (1 - 4) Left | 24.5 | 28.0 | 25.3 |
|  | 2 (5 - 6) Centre | 29.2 | 35.0 | 37.3 |
|  | 3 (7 -10) Right | 21.5 | 18.2 | 19.6 |

|  | | Skeptical (%) | Confident (%) | Tradeoff (%) |
|---|---|---|---|---|
|  | 9 DK/Refusal | 24.8 | 18.8 | 17.8 |
| Life satisfaction | 1 Very satisfied | 18.2 | 28.0 | 25.7 |
|  | 2 Fairly satisfied | 54.4 | 57.4 | 57.0 |
|  | 3 Not very satisfied | 20.1 | 11.3 | 13.8 |
|  | 4 Not at all satisfied | 6.9 | 2.8 | 3.4 |
|  | 5 DK | 0.4 | 0.5 | 0.2 |
| Total |  | 100 | 100 | 100 |

Chi-Square tests of association statistically significant for p=0.001

## Conclusions

The 2019 Eurobarometer 91.2 survey of the EU27-UK population indicates that there are large segments of the public who believe in probable vaccine damage, either in general or specifically related to immune system disorders or causing the illness they are meant to prevent. With substantial variation at the country level, we found proportions as high as 64% in Croatia and 60% in France, or as low as 26% in Sweden, who considered that "Vaccines can often produce serious side-effects," with an EU27-UK average of almost half subscribing to this belief (48%). Therefore, we can consider that "probable vaccine damage" has become an alternative fact, co-generated by people who offer and seek information from online and offline sources in a diverse social space of vaccine knowledge, experience, and trust in science and medical authority.

Beliefs in probable vaccine damage are not limited to people who mistrust vaccines. We identified three patterns of beliefs as regards vaccine effectiveness, safety, and usefulness: the skeptic, the confident, and the trade-off types. While the skeptics are relatively low on vaccine trust in all three dimensions, and the confident are relatively high, the trade-off type displays a combination of beliefs in probable vaccine damage (higher than the skeptics, on average) with trust in vaccine effectiveness, testing, and usefulness. The confident and the trade-off types have largely similar professed vaccination histories, with superior rates for the confident, while the skeptical have much lower rates of vaccination for themselves and for their children. Moreover, there are no strong sociodemographic differences among the three belief clusters, though the skeptical are more likely to position themselves on the right side of an ideological continuum and to be less satisfied with life.

The concomitance of beliefs in probable vaccine damage and effectiveness, testing, and usefulness of vaccines has implications for designing public communication campaigns. In particular, campaigns that focus on the value of vaccines in preventing disease will not address the subjective reality of vaccination damage, which is shared by large segments of the public that otherwise trust vaccination. For example, Romania is one of the countries with a high proportion of trade-off respondents. Romania has also suffered from an ongoing measles epidemic since 2016, which has infected more than 19,500 people and led to 64 deaths as reported on April 3, 2020[1]. The Romanian state has conducted a media campaign under the slogan "Vaccination saves lives" including three videos that focus on the effectiveness and usefulness of vaccination against disease, in history and today[2]. Still, in light of the present findings, it becomes clear that such a campaign, which is aimed at the skeptics (about 23% in Romania), does not address the trade-off segment of the public, comprising about 25% of Romanian respondents. In Romania's campaign and in global vaccination campaigns, new patterns of public communication about vaccination are needed to address the alternative fact of probable vaccine damage and to consider the trade-off patterns of concomitant trust and mistrust in assessing vaccines.

---

[1] National Institute for Public Health in Romania, Situation of Measles in Romania. April 3, 2020 Report. Available online at https://www.cnscbt.ro/index.php/informari-saptamanale/rujeola-1/1622-situatia-rujeolei-in-romania-la-data-de-03-04-2020/file

[2] Videos are available on the YouTube channel of the Romanian Ministry of Health, under the titles: "Vaccinurile salvează vieți - Paula Rusu" (about tuberculosis), "Vaccinurile salvează vieți - Victor Rebengiuc" (about the history of vaccination and polio), and "Vaccinurile salvează vieți - Ioana Grozea" (about childhood vaccination in general). Available online at https://www.youtube.com/channel/UCAR1cIye4_xGGZoBZvvjMQQ

## Author contribution

All authors made a significant contribution to the development of this manuscript and approved the final version for submission.

## Funding

This work is part of the research project PN-III-P4-ID-PCE-2020-1589 "The manufacture of doubt on vaccination and climate change. A comparative study of legitimacy tactics in two science-skeptical discourses", funded by the Ministry of Education and Scientific Research.

## Declaration of competing interests

The authors declare that they have no known competing financial interests or personal relationships that could have appeared to influence the work reported in this paper.

## Appendix A. Supplementary material

Table 4. Country distribution of belief configurations regarding vaccine effectiveness, safety and usefulness. Data source: Eurobarometer 91.2

## Appendix B. Supplementary material

Table 5. Country distribution of beliefs in vaccine effectiveness, risks and usefulness. Data source: Eurobarometer 91.2

**Appendix A. Supplementary material**

*Table 4. Country distribution of belief configurations regarding vaccine effectiveness, safety and usefulness. Data source: Eurobarometer 91.2*

|  | Skeptical | Confident | Tradeoff | Total |
|---|---|---|---|---|
|  | % | % | % | % |
| France | 18 | 49 | 32 | 100 |
| Belgium | 11 | 54 | 36 | 100 |
| The Netherlands | 3 | 77 | 20 | 100 |
| Germany | 8 | 58 | 34 | 100 |
| Italy | 20 | 55 | 25 | 100 |
| Luxembourg | 13 | 57 | 30 | 100 |
| Denmark | 6 | 69 | 25 | 100 |
| Ireland | 10 | 57 | 32 | 100 |
| United Kingdom | 6 | 65 | 29 | 100 |
| Greece | 9 | 66 | 25 | 100 |
| Spain | 5 | 67 | 28 | 100 |
| Portugal | 2 | 66 | 32 | 100 |
| Finland | 4 | 68 | 28 | 100 |
| Sweden | 5 | 78 | 17 | 100 |
| Austria | 24 | 49 | 27 | 100 |
| Cyprus (Republic) | 6 | 53 | 40 | 100 |
| Czech Republic | 9 | 51 | 40 | 100 |
| Estonia | 9 | 66 | 24 | 100 |
| Hungary | 13 | 55 | 32 | 100 |
| Latvia | 21 | 50 | 29 | 100 |
| Lithuania | 12 | 55 | 33 | 100 |
| Malta | 4 | 46 | 50 | 100 |
| Poland | 8 | 66 | 26 | 100 |
| Slovakia | 9 | 57 | 34 | 100 |
| Slovenia | 11 | 46 | 44 | 100 |
| Bulgaria | 13 | 67 | 20 | 100 |
| Romania | 23 | 52 | 25 | 100 |
| Croatia | 18 | 44 | 38 | 100 |
| **TOTAL EU27-UK** | **11** | **59** | **29** | **100** |

Chi-Square tests of association statistically significant for p=0.001

**Appendix B. Supplementary material**

*Table 5. Country distribution of beliefs in vaccine effectiveness, risks and usefulness. Data source: Eurobarometer 91.2*

| | Are vaccines effective | Vaccines overload and weaken the immune system | Vaccines can cause the disease against which they protect | Vaccines can often produce serious side effects | Vaccines are not rigorously tested before being authorized for use | It is important for everybody to have routine vaccinations | Vaccines are only important for children | Not getting vaccinated can lead to serious health issues | Vaccines are important to protect not only yourself but also others | Vaccination of other people is important to protect those that cannot be vaccinated |
|---|---|---|---|---|---|---|---|---|---|---|
| | % | % | % | % | % | % | % | % | % | % |
| **Total – EU27 and the UK** | **87** | **31** | **38** | **48** | **11** | **85** | **29** | **85** | **91** | **91** |
| Austria | 75 | 42 | 39 | 51 | 12 | 76 | 38 | 76 | 78 | 81 |
| Belgium | 81 | 34 | 48 | 53 | 9 | 83 | 33 | 83 | 94 | 92 |
| Bulgaria | 79 | 25 | 25 | 41 | 7 | 89 | 45 | 80 | 88 | 87 |
| Croatia | 81 | 45 | 43 | 64 | 18 | 84 | 45 | 80 | 84 | 86 |
| Cyprus (Republic) | 85 | 39 | 42 | 65 | 8 | 93 | 21 | 87 | 93 | 94 |
| Czech Republic | 91 | 46 | 45 | 53 | 8 | 93 | 32 | 79 | 90 | 91 |
| Denmark | 94 | 21 | 42 | 35 | 8 | 91 | 9 | 92 | 96 | 92 |
| Estonia | 88 | 30 | 32 | 41 | 8 | 88 | 26 | 86 | 93 | 94 |
| Finland | 97 | 16 | 44 | 41 | 11 | 95 | 14 | 95 | 98 | 97 |
| France | 84 | 34 | 45 | 60 | 12 | 70 | 18 | 76 | 89 | 89 |

| | Are vaccines effective | Vaccines overload and weaken the immune system | Vaccines can cause the disease against which they protect | Vaccines can often produce serious side effects | Vaccines are not rigorously tested before being authorized for use | It is important for everybody to have routine vaccinations | Vaccines are only important for children | Not getting vaccinated can lead to serious health issues | Vaccines are important to protect not only yourself but also others | Vaccination of other people is important to protect those that cannot be vaccinated |
|---|---|---|---|---|---|---|---|---|---|---|
| Germany | 89 | 33 | 42 | 46 | 8 | 88 | 20 | 88 | 91 | 92 |
| Greece | 90 | 29 | 30 | 53 | 9 | 86 | 33 | 84 | 90 | 93 |
| Hungary | 82 | 39 | 38 | 44 | 16 | 88 | 48 | 85 | 88 | 89 |
| Ireland | 89 | 34 | 35 | 57 | 11 | 82 | 37 | 86 | 91 | 92 |
| Italy | 79 | 31 | 34 | 46 | 18 | 73 | 50 | 79 | 83 | 84 |
| Latvia | 71 | 39 | 39 | 55 | 12 | 76 | 33 | 71 | 87 | 88 |
| Lithuania | 84 | 31 | 44 | 55 | 8 | 89 | 38 | 78 | 91 | 92 |
| Luxembourg | 88 | 30 | 38 | 53 | 9 | 80 | 18 | 82 | 93 | 88 |
| Malta | 97 | 43 | 51 | 62 | 1 | 91 | 17 | 94 | 97 | 95 |
| Poland | 85 | 28 | 30 | 40 | 13 | 94 | 53 | 91 | 91 | 89 |
| Portugal | 93 | 28 | 34 | 50 | 4 | 97 | 21 | 97 | 99 | 98 |
| Romania | 76 | 33 | 36 | 53 | 20 | 77 | 52 | 76 | 79 | 81 |
| Slovakia | 87 | 37 | 37 | 49 | 7 | 90 | 41 | 88 | 89 | 91 |
| Slovenia | 85 | 49 | 47 | 60 | 7 | 87 | 34 | 83 | 91 | 90 |
| Spain | 95 | 28 | 37 | 43 | 9 | 93 | 16 | 85 | 96 | 93 |
| Sweden | 96 | 15 | 33 | 26 | 7 | 88 | 12 | 91 | 96 | 97 |

|  | Are vaccines effective | Vaccines overload and weaken the immune system | Vaccines can cause the disease against which they protect | Vaccines can often produce serious side effects | Vaccines are not rigorously tested before being authorized for use | It is important for everybody to have routine vaccinations | Vaccines are only important for children | Not getting vaccinated can lead to serious health issues | Vaccines are important to protect not only yourself but also others | Vaccination of other people is important to protect those that cannot be vaccinated |
|---|---|---|---|---|---|---|---|---|---|---|
| The Netherlands | 98 | 15 | 38 | 29 | 4 | 95 | 10 | 92 | 97 | 96 |
| United Kingdom | 92 | 27 | 33 | 54 | 7 | 89 | 22 | 90 | 96 | 95 |

Chi-Square tests of association statistically significant for p=0.001